\documentclass[]{andp2012}
\usepackage{graphicx}
\usepackage[ansinew]{inputenc}
\usepackage{array}
\usepackage{color}
\usepackage{amsmath}
\usepackage{amsxtra}
\usepackage{amstext}
\usepackage{amssymb}
\usepackage{latexsym}
\usepackage{dsfont}
\usepackage{verbatim}

\begin{document}

\title{A short walk through quantum optomechanics}

\author{P. Meystre}
\date{\today}

\begin{abstract}
This paper gives an brief review of the basic physics of quantum optomechanics and provides an overview of some of its recent developments and current areas of focus. It first outlines the basic theory of cavity optomechanical cooling and gives a brief status report of the experimental state-of-the-art. It then turns to the deep quantum regime of operation of optomechanical oscillators and cover selected aspects of quantum state preparation, control and characterization, including mechanical squeezing and pulsed optomechanics. This is followed by a discussion of the ``bottom-up'' approach that exploits ultracold atomic samples instead of nanoscale systems. It concludes with an outlook that concentrates largely on the functionalization of quantum optomechanical systems and their promise in metrology applications.
\end{abstract}

\maketitle

\section{Introduction}

Broadly speaking, quantum optomechanics provides a universal tool to achieve the quantum control of mechanical motion~\cite{reviews}. It does that in devices spanning a vast range of parameters, with mechanical frequencies from a few Hertz to GHz, and with masses from $10^{-20}$g to several kilos. At a fundamental level, it offers a route to determine and control the quantum state of truly macroscopic objects and paves the way to experiments that may lead to a more profound understanding of quantum mechanics; and from the point of view of applications, quantum optomechanical techniques in both the optical and microwave regimes will provide motion and force detection near the fundamental limit imposed by quantum mechanics.

While many of the underlying ideas of quantum optomechanics can be traced back to the study of gravitational wave detectors in the 1970s and 1980s \cite{BraginskyCaves,MeystreScully}, the spectacular developments of the last few years rely largely on two additional developments: From the top down, it is the availability of advanced micromechanical and nanomechanical devices capable of probing extremely tiny forces, often with spatial resolution at the atomic scale.  And from the bottom-up, we have gained a detailed understanding of the mechanical effects of light and how they can be exploited in laser trapping and cooling. These developments open a path to the realization of macroscopic mechanical systems that operate deep in the quantum regime, with no significant thermal noise remaining. As a result, they offer both knowledge and control of the quantum state of a macroscopic object, and increased sensitivity, precision, and accuracy in the measurement of feeble forces and fields.

It was Arthur Ashkin \cite{Ashkin} who first suggested and demonstrated that small dielectric balls can be accelerated and trapped using the radiation-pressure forces associated with focused laser beams. In later experiments these particles, weighting on the order of a microgram, were levitated against the Earth gravitational field. This advance led to the realization of optical tweezers, whose applications in biological science have become ubiquitous. In parallel, the use of the strong enhancement provided by resonant light scattering lead to the laser cooling of ions and of neutral atoms by D. Wineland, T. W. H{\"a}nsch, S. Chu, W. D. Phillips, C. Cohen-Tannoudji and many others, resulting in a wealth of extraordinary developments~\cite{Metcalf} culminating in 1995 with the invention of atomic Bose-Einstein condensates~\cite{Cornell95,Ketterle95}, and the subsequent explosion in the study of quantum-degenerate atomic systems.

Non-resonant light-matter interactions present the considerable advantage of being largely wavelength independent, providing one with the potential to achieve optomechanical effects for a broad range of wavelengths from the microwave to the optical regime. Resonant interactions, on the other hand, can result in a very large enhancement of the interaction, but at the cost of being limited to narrow ranges of wavelengths. Cavity optomechanics exploits the best of both worlds by achieving resonant enhancement through an engineered resonant structure rather than via the internal structure of materials. This could be for example an optical resonator with a series of narrow resonances, or an electromagnetic resonator such as a superconducting LC circuit. Indeed, numerous designs can achieve optomechanical control via radiation pressure effects in high-quality resonators. They range from nanometer-sized devices of as little as $10^7$ atoms to micromechanical structures of $10^{14}$ atoms and to macroscopic centimeter-sized mirrors used in gravitational wave detectors. 

That development first appeared at the horizon in the 1960s, but more so in the late 1970s and 1980s. It was initially largely driven by the developments in optical gravitational wave antennas spearheaded by V. Braginsky, K. Thorne, C. Caves, and others~\cite{BraginskyCaves,BraginskyBook,MeystreScully}. These antennas operate by coupling kilogram-size test masses to the end-mirrors of a large path length optical interferometer.  Changes in the optical path length due to local changes in the curvature of space-time modulate the frequency of the cavity resonances and in turn, modulate the optical transmission through the interferometer.  It is in this context that researchers understood fundamental quantum optical effects on mechanics and mechanical detection such as the standard quantum limit, and how the basic light-matter interaction can generate non-classical states of light. 

Braginsky and colleagues demonstrated cavity optomechanical effects with microwaves \cite{Braginsky67} as early as 1967. In the optical regime, the first demonstration of these effects was the radiation-pressure induced optical bistability in the transmission of a Fabry-P\'erot interferometer, realized by Dorsel {\it el al.} in 1983~\cite{Dorsel}. In addition to these adiabatic effects, cooling or heating of the mechanical motion is also possible,  due to the finite time delay between the mechanical motion and the response of the intracavity field, see section 2.2. The cooling effect was first observed in the microwave domain by Blair {\it et al.}~\cite{Blair} in a Niobium high-$Q$ resonant mass gravitational radiation antenna, and 10 years later in the optical domain in several laboratories around the world: first via feedback cooling of a mechanical mirror by Cohadon {\it et al.}~\cite{Cohadon99}, followed by photothermal cooling by Karrai and coworkers~\cite{Hohberger04}, and shortly thereafter by radiation pressure cooling in several groups~\cite{Gigan06,Arcizet06,Kleckner06,Schliesser06,Corbitt07,Vinante08}. Also worth mentioning is that as early as 1998  Ritsch and coworkers proposed a related scheme to cool atoms inside a cavity \cite{Hechenblaikner98}. 

This paper reviews the basic physics of quantum optomechanics and gives a brief overview of some of its recent developments and current areas of focus. Section 2 outlines the basic theory of cavity optomechanical cooling and sketches a brief status report of the experimental state-of-the-art in ground state cooling of mechanical oscillators,  a snapshot of a situation likely to be rapidly outdated.  Of course ground state cooling is only the first step in quantum optomechanics. Quantum state preparation, control and characterization are the next challenges of the field. Section 3 gives an overview of some of the major trends in this area, and discusses topics of much current interest such as the so-called strong-coupling regime, mechanical squeezing, and pulsed optomechanics. Section 4 discusses a complementary ``bottom-up'' approach that exploits ultracold atomic samples instead of nanoscale systems to study quantum optomechanical effects. Finally, Section 5 is an outlook that concentrates largely on the functionalization of quantum optomechanical systems and their promise in metrology applications.

\section{Basic theory}
To describe the basic physics underlying the main aspects of cavity optomechanics it is sufficient to consider an optically driven Fabry-P{\'e}rot resonator with one end mirror fixed -and effectively assumed to be infinitely massive, and the other harmonically bound and allowed to oscillate under the action of radiation pressure from the intracavity light field of frequency $\omega_L$, see Fig.~1. Braginsky recognized as early as 1967~\cite{Braginsky67} that as radiation pressure drives the mirror, it changes the cavity length, and hence the intracavity light field intensity and phase. This results in two main effects: the ''optical spring effect,'' an optically induced change in the oscillation frequency of the mirror that can produce a significant stiffening of its effective frequency; and optical damping, or ``cold damping,'' whereby the optical field acts effectively as a viscous fluid that can damp the mirror motion and cool its center-of-mass motion. 

\begin{figure}[t]
\includegraphics[width=0.45\textwidth]{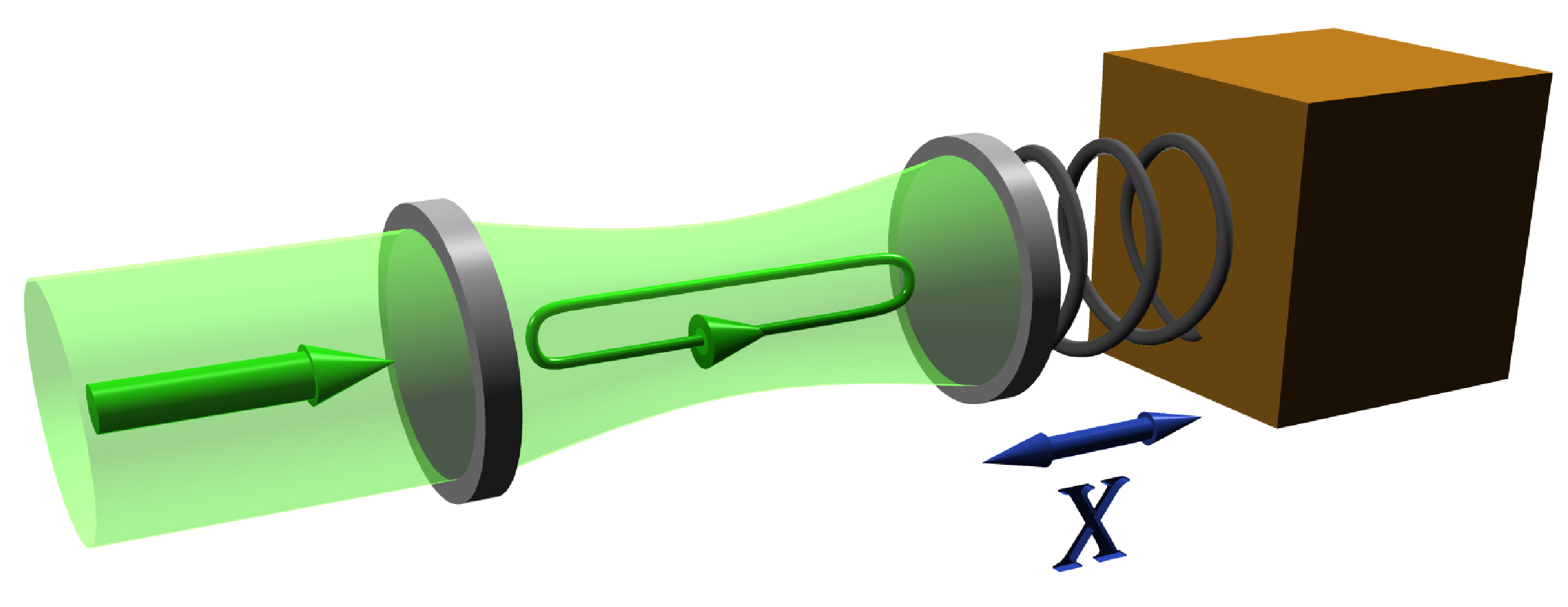}
\caption{(Color online) Generic cavity optomechanical system. The cavity consists of a highly reflective fixed input mirror and a small movable end mirror harmonically coupled to a support that acts as a thermal reservoir.  }
\end{figure}

One can immediately understand how the optical spring effect  can result in a more quantum behavior of the oscillator by recalling that in the high temperature limit the mean number of phonons $\langle n_m \rangle$ in the center-of-mass motion of an oscillator of frequency $\Omega_m$ is given by
\begin{equation}
\langle n_m\rangle =k_B T/\hbar \Omega_m,
\end{equation}
where $k_B$ is Boltzman's constant and $T$ the temperature. For a given temperature, increasing $\Omega_m$ automatically reduces $\langle n_m\rangle$, allowing one to approach the quantum regime without having to reduce the temperature.

Cold damping, in contrast, does reduce the temperature of the oscillating mirror by opening up a dissipation channel to a reservoir that is effectively at zero temperature. To see how this work, we first remark that in the absence of optical field the oscillating mirror is dissipatively coupled to a thermal bath at temperature $T$.  Its average center-of-mass energy, $\langle E \rangle$, results from the balance between dissipation and heating,
\begin{equation}
\frac{d\langle E\rangle}{dt}= -\gamma \langle E \rangle + \gamma k_BT,
\end{equation}
where $\gamma$ is the intrinsic mechanical damping rate. When an optical field is applied, an additional optomechanical damping channel with damping rate $\Gamma_{\rm opt}$ comes into play so that
\begin{equation}
\frac{d\langle E\rangle}{dt}= -\gamma \langle E \rangle + \gamma k_BT -\Gamma_{\rm opt} \langle E \rangle.
\label{balance}
\end{equation}
Importantly, that channel does not come with an additional (classical) thermal bath.  Optical frequencies are much higher than mechanical frequencies, so that the optical field is effectively coupled to a reservoir at zero temperature. In steady state Eq.~(\ref{balance}) gives $\langle E\rangle = k_B T/(\gamma + \Gamma_{\rm opt})$, or
\begin{equation}
T_{\rm eff}= \frac{\gamma T}{\gamma + \Gamma_{\rm opt}}.
\label{cold-damping}
\end{equation}
This simple phenomenological  classical picture predicts that the fundamental limit of cooling is $T=0$. A more detailed quantum mechanical analysis does yield a fundamental limit given by quantum noise, see Section IIC, but in practice, this is usually not a major limitation to cooling the mechanical mode arbitrarily close to the quantum ground state, $\langle n_m \rangle =0$.

More quantitatively, we consider a single mode of the optical resonator of nominal frequency $\omega_c$ and assume that radiation pressure results in a displacement $x(t)$ of the harmonically bound end-mirror, and consequently in a change in the optical mode resonance frequency to
\begin{equation}
\omega_c'=\omega_c - G x(t),
\label{omegax}
\end{equation}
where 
\begin{equation}
G=-\partial \omega_c'/\partial x.
\label{G}
\end{equation}
For a single-mode Fabry-P{\'e}rot resonator of length $L$ this becomes simply $G=\omega_c/L$.
 
Typical mechanical oscillator frequencies are in the range of $\Omega_m/2\pi=$10Hz to $10^9$Hz and the mechanical quality factors of the mirrors are in the range of perhaps $Q_m \approx 10^3-10^7$, so that typically the damping rate $\Gamma=\Omega_m/Q_m$ of the oscillating mirror is much slower than the damping rate $\kappa$ of the intracavity field. One can then gain considerable intuition by first neglecting mirror damping altogether and assuming that its motion is approximately  harmonic, 
\begin{equation}
x(t) \approx x_0 \sin(\Omega_m t).
\end{equation}
For a classical monochromatic pump of frequency $\omega_L$ and amplitude $\alpha_{\rm in}$ the intracavity field obeys the equation of motion
\begin{equation}
\frac{d \alpha(t)}{dt} = \left [ i\left (\Delta + G x(t) \right ) - \kappa/2 \right ]\alpha(t) + \sqrt{\kappa} \alpha_{\rm in},
\label{field eq}
\end{equation}
with the steady-state solution
\begin{equation}
\alpha=\frac{\sqrt{\kappa}\alpha_{\rm in}}{-i(\Delta + G x) + \kappa/2}.
\end{equation}
Here we have introduced the detuning 
\begin{equation}
\Delta = \omega_L-\omega_c
\end{equation} 
and $\alpha$ is the intracavity field amplitude, normalized so that
\begin{eqnarray}
|\alpha|^2&=&\frac{\kappa}{(\Delta + G x)^2 + (\kappa/2)^2}\left (\frac{P}{\hbar \omega_L}\right ) \nonumber \\
&=&\frac{\kappa}{(\Delta + \omega_cx/L)^2 + (\kappa/2)^2}\left (\frac{P}{\hbar \omega_L} \right )
\label{steadystate}
\end{eqnarray}
where 
\begin{equation}
P=\hbar \omega_L |\alpha_{\rm in}|^2
\end{equation}
 is the input laser power driving the cavity mode. This normalization allows for an easy generalization to the case of quantized fields, in which case $\alpha$ will be interpreted as the square root of the mean number of intracavity photons, $\alpha =\sqrt {\langle \hat a^\dagger \hat a \rangle}$, with $\hat a$ and $\hat a^\dagger$ the bosonic annihilation and creation operators of the intracavity field. Note that $|\alpha_{\rm in}|^2$ has then the units of ``photons per second.'' 

For oscillation amplitudes $x_0$ small enough that  $G x_0/\Omega_m\ll 1$, it can be shown that the mirror oscillations simply result in the generation of two sidebands at frequencies $\omega_L \pm \Omega_m$, see e.g. Refs.~\cite{SchliesserThesis,KippenbergOE}. The time-dependent complex field amplitude $\alpha(t)$ then takes the approximate form $\alpha(t) \simeq \alpha_0(t) + \alpha_1(t)$
with
\begin{eqnarray}
\alpha_0(t) &\simeq& \frac{\sqrt{\kappa} \alpha_{\rm in}}{-i\Delta + \kappa/2}, \\
\alpha_1(t)&\simeq&\left (\frac{G x_0}{2}  \right )\frac{\sqrt{\kappa} \alpha_{\rm in}}{-i\Delta + \kappa/2}  \\
&\times& \left (\frac{e^{-i \Omega_m t}}{-i(\Delta + \Omega_m) + \kappa/2} - \frac{e^{+i\Omega_m t}}{-i(\Delta - \Omega_m) + \kappa/2}\right ).\nonumber
\label{sideband}
\end{eqnarray}
The first sideband in Eq.~(\ref{sideband}) can be interpreted as an anti-Stokes line, with a resonance at $\omega_L=\omega_c-\Omega_m$, and the second one is a Stokes line. An important feature of these sidebands is that their amplitudes can be vastly different, as they are determined by the cavity Lorentzian response function evaluated at $\omega_L-\Omega_m$ and $\omega_L+\Omega_m$, respectively.

\subsection{Static phenomena}

Consider first a situation where the cavity damping rate $\kappa$ is much faster than all other characteristic times of the system. One can then understand the mirror motion as resulting from the combined effects of the harmonic restoring force and the radiation pressure force $F_{\rm rp}$ resulting from an adiabatic elimination of the intracavity field, see e.g. Ref.~\cite{MarquardtLesHouches},
\begin{equation}
 F_{\rm rp}= \hbar G |\alpha|^2 = \hbar \frac{\omega_c}{L} |\alpha|^2,
\label{classical opto-coupling}
 \end{equation}
 where $|\alpha|^2$ is given by Eq.~(\ref{steadystate}) and the second equality holds for a simple Fabry-P{\'e}rot. One can easily show that the force $F_{\rm rp}$ can be derived from the potential 
\begin{equation}
V_{\rm rp}=-\frac{\hbar \kappa |\alpha|^2 }{2} \arctan \left[2(\Delta + Gx)/\kappa \right ],
\end{equation}
the mirror of mass $m$ being therefore subject to the total potential
\begin{equation}
V(x) =\frac{1}{2} m \Omega_m^2 x^2  -\frac{\hbar \kappa |\alpha|^2 }{2} \arctan \left[2(\Delta + G x)/\kappa \right ].
\end{equation} 
The potential $V_{\rm rp}$ slightly displaces the equilibrium position of the mirror to a position $x_0\neq 0$, as would be intuitively expected, and also changes its spring constant from its intrinsic value $k=m \Omega_m^2$ to a new value
\begin{equation}
k_{\rm rp}=m \Omega_m^2 +\left .  \frac{d^2 V_{\rm rp}(x)}{dx^2} \right |_{x=x_0}.
\end{equation}
The second term in this expression is the static {\em optical spring effect}. For realistic parameters it can increase the stiffness of the mechanical system by orders of magnitude. A third important static effect of radiation pressure is that in general, there is a range of parameters for which the potential $V(x)$ can exhibit 3 extrema. Two of them correspond to stable local minima of $V(x)$, and the third one to an unstable maximum. This results in radiation pressure induced optical bistability~\cite{Dorsel}, an effect that  is physically similar to the more familiar form of bistability that can occur in a Kerr nonlinear medium. The difference is that in one case, it is the optical length of the resonator that is changed by a Kerr nonlinearity, with its physical length remaining unchanged, while in the other it is that physical length that is intensity-dependent.
 
\subsection{Effects of retardation} 

In general the optical field does not respond instantly to the motion of the mechanical oscillator, therefore we need to account for the effects of retardation as well. We proceed by assuming that he system is in equilibrium at some mirror position $x_0$ with intracavity field $\alpha_0$, taken to be real without loss of generality, and consider the linearized dynamics of small displacements  $\delta x(t)$ and $\delta \alpha(t)$ from that state under the effect of an external force $\delta F(t)$,
\begin{eqnarray}
\ddot {\delta x} + \Gamma \dot {\delta x} + \Omega_m^2 \delta x &=& \hbar G \alpha_0\left (\delta \alpha + \delta \alpha^* \right ), \nonumber \\
\dot{\delta \alpha} &=& (i\Delta -\kappa/2)\delta \alpha + i G  \alpha_0 \delta x.
\label{linearized}
\end{eqnarray}
These equations of motion can easily be solved, for instance in Fourier space, to give
\begin{equation}
\delta \alpha (\omega)= \left ( \frac{iG \alpha_0}{-i(\bar \Delta + \omega) + \kappa/2}\right )   \delta x(\omega)
\label{delta alfa}
\end{equation}
where 
\begin{equation}
\bar {\Delta} = \Delta + G x_0,
\end{equation}
resulting in a modification of the radiation pressure force
\begin{equation}
\delta F_{\rm rp}(\omega)=-\hbar G \alpha_0 \left [ \delta \alpha(\omega) + \delta \alpha^*(\omega) \right ].
\end{equation}
Together with Eq.~(\ref{delta alfa}) this expression shows that the mirror motion exerts a {\em dynamical back-action} on the radiation pressure force, which acquires both a real and an imaginary component, the physical origin of the imaginary component being the delayed response of the intracavity field. As a result the intracavity power acquires a component that oscillates out of phase with the mirror motion, that is, with its velocity. It is through that friction force that the optical field acts as a viscous field for the mirror. 

The net effect of the real and imaginary components of $\delta F_{\rm rp}$ can be conveniently cast in terms of the back-action frequency shift $\delta \Omega_{\rm opt}$ and a damping rate $\Gamma_{\rm opt}$, see e.g. Ref.~\cite{SchliesserThesis,KippenbergOE}, with
\begin{equation}
\delta \Omega_{\rm opt}=\frac{\hbar G^2 \alpha_0^2}{2m\Omega_m} \left [ \frac{{\bar \Delta}+\Omega_m}{(\bar {\Delta} + \Omega_m)^2+\kappa^2/4} + \frac{{\bar \Delta}+\Omega_m}{(\bar {\Delta} - \Omega_m)^2+\kappa^2/4} \right ]
\end{equation}
and
\begin{equation}
\Gamma_{\rm opt}=\frac{\hbar G^2 \alpha_0^2}{2m\Omega_m}\left [ \frac{\kappa}{(\bar {\Delta} + \Omega_m)^2+\kappa^2/4} - \frac{\kappa}{(\bar {\Delta} - \Omega_m)^2+\kappa^2/4}  \right ].
\label{gamma-ba}
\end{equation}
For detunings $\bar \Delta \approx -\Omega_m$ the first term in Eq.~(\ref{gamma-ba}) dominates over the second term, and the dynamical back-action results in an increase in the mechanical damping of the mechanical oscillator and cooling, see Eq.~(\ref{cold-damping}). It is therefore the asymmetry between the response function of the Fabry-P\'erot at the frequencies of the two side modes that is responsible for cooling -- or ``anti-damping''  if one changes the sign of $\Delta$ and uses a blue-detuned instead of a red-detuned driving field. In particular, in the {\em resolved sideband limit} $\Omega_m \gg \kappa$ we find
\begin{equation}
\Gamma_{\rm opt} \approx\left ( \frac {2}{\kappa} \right ) \frac{\hbar G^2  \alpha_0^2}{m\Omega_m} ,
\end{equation}
which can in principle be increased arbitrarily (within the limits of validity of the model) by increasing the incident optical power. 

Together with Eq.~(\ref{cold-damping}) this analysis predicts that the cooling of the center-of-mass motion of the mirror can be arbitrarily close to $T_{\rm eff}=0$, a consequence of the fact that the optomechanical coupling between the intracavity field and the mirror results in the scattering of the the driving field into an anti-Stokes line that is strongly damped due to the high density of states at the cavity resonance.  Conversely, for the opposite detuning $\bar \Delta \approx -\Omega_m$ it is the Stokes line that is strongly damped, resulting in anti-damping of the mirror motion.  This can lead to parametric oscillations and dynamical instabilities, a situation further discussed in section 3.5.

\begin{figure}[t]
\includegraphics[width=0.45\textwidth]{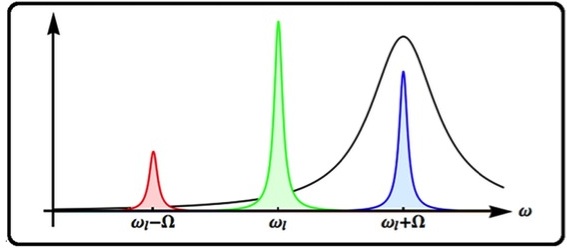}
\caption{(Color online) Schematic of sideband cooling: a coherent light field driving the resonator acquires frequency sidebands due to the mirror oscillations. The origin of the high frequency sideband is the parametric transfer of phonons from the mirror to the optical field and the lower sideband is due to the reverse process, see section 2.2. Sideband cooling results when the upper sideband frequency is resonant with the resonator. The solid black curve depicts the resonator transmission near its mode of frequency $\omega_c$. }
\end{figure}
The quantum description of the next section will show that cold damping and mirror cooling can also be interpreted in terms of of the annihilation of phonons from the center-of-mass mode of oscillation when scattering the driving laser field into the anti-Stokes sideband. Heating can similarly be understood as resulting from the creation of phonons associated with the scattering of the driving field into the lower frequency Stokes side mode.

\subsection{Quantum limit}
\label{tql}
The classical prediction that one can in principle reach an arbitrarily large degree of cooling needs to be revised to account for the effects of  quantum and thermal noise. As is well known, the open port of the interferometer used to supply the optical drive of the oscillating mirror also allows for the coupling of vacuum fluctuations into the resonator, see e.g. Ref.~\cite{Caves}. This leads to a fundamental limit to the degree of cooling that can be achieved. A proper quantum description of the system must account for this effect as well as for the the bosonic nature of the phonons. 

Ignoring in a first step the important effects of fluctuations and dissipation, and in case a single optical mode of the Fabry-P{\'e}rot resonator and a single mode of oscillation of the suspended mirror need to be considered,  the optomechanical Hamiltonian is simply
\begin{equation}
H=\hbar \omega(\hat x) \hat a^\dagger \hat a + \frac{\hat p^2}{2m} + \frac{1}{2} m \Omega_m^2 \hat x^2,
\label{starting H}
\end{equation}
where $\hat a$ and $\hat a^\dagger$ are bosonic annihilation and creation operators for the cavity mode of frequency $\omega$, and $\hat p$ and $\hat x$ are the momentum and position of the oscillating mirror of mass $m$ and frequency $\Omega_m$. In reality, though, this Hamiltonian is more subtle than may appear at first. This is because the mode frequency $\omega(q)$ depends on the length of the resonator, which in turn depends on the intracavity intensity. Stated differently, the boundary conditions for the quantization of the light field are changing in time, and do so in a fashion that depends on the state of that field and its history. The rigorous quantization of this system is a far-from-trivial problem, but for most cases of interest in quantum optomechanics the situation is significantly simplified since the transit time $c/2L$ of the light field through the optical resonator is much faster than the mechanical frequency $\Omega_m$. The intracavity field therefore ``learns'' about changes in its environment in times short compared to $1/\Omega_m$. Under these conditions one can assume that the cavity frequency follows adiabatically any change in resonator length,
\begin{equation}
\omega(\hat x)= \frac{n\pi c}{L+\hat x}=\omega_c\left (\frac{1}{ 1+ \hat x/L} \right ) \approx \omega_c(1-\hat x/L)
\end{equation}
where $n$ is an integer that labels the mode of nominal frequency $\omega_c$  and $L$ is the nominal resonator length (in the absence of light.) In the classical limit we recover the result $G= \omega_c/L$ valid for a simple Fabry-P\'erot. The Hamiltonian (\ref{starting H}) reduces then to~\cite{Law95,Mancini,Knight}
\begin{eqnarray}
\label{H quantum opto}
H&=&\hbar \omega_c \hat a^\dagger \hat a + \frac{\hat p^2}{2m} + \frac{1}{2} m \Omega_m^2 \hat x^2 -\hbar G \hat a^\dagger \hat a \hat x \label{H opto} \\
&=&\hbar \omega_c \hat a^\dagger \hat a + \frac{\hat p^2}{2m} + \frac{1}{2} m \Omega_m^2 \hat x^2 -\hbar g_0 \hat a^\dagger \hat a (\hat b + \hat b^\dagger).\nonumber
\end{eqnarray}
In the second line we have used the familiar relationship between the position operator $\hat x$ and the annihilation and creation operators $\hat b$ and $\hat b^\dagger$ of the mechanical oscillator,
\begin{equation}
\hat x = x_{\rm zpt} (\hat b + \hat b^\dagger)
\end{equation}
with
\begin{equation}
x_{\rm zpt} = \sqrt{\frac{\hbar}{2m \Omega_m}}.
\end{equation}
We also introduced the optomechanical coupling frequency
\begin{equation}
g_0 = x_{\rm zpf} G = -x_{\rm zpf} \partial \omega_c'/\partial x,
\end{equation}
which scales  the optomechanical displacement to the zero-point motion of the mechanical oscillator. The Hamiltonian (\ref{H opto}) is the starting point for most quantum mechanical discussions of cavity optomechanics.

In order to establish the theoretical limit to cavity optomechanical cooling, it is necessary to expand the description provided by the Hamiltonian~(\ref{H opto}) to account for the optical drive of the resonator, cavity damping, and the mechanical damping of the oscillator. This analysis was carried out in Refs.~\cite{Marquardt07,WilsonRae07,Marquardt08}. The main message of these papers is that -- at least for constant optomechanical coupling -- the best cooling can be achieved in the so-called resolved sideband limit, $\kappa \ll \Omega_m$, with the minimum mean phonon number
\begin{equation}
\langle n_m \rangle = \frac{\Gamma_{\rm opt} \bar n_m^0 + \gamma \bar n_m^T}{\gamma + \Gamma_{\rm opt}}.
\end{equation}
Here
$\bar n_m^0$ is the mean steady-state number of phonons in the absence of mechanical damping, given by the detailed balance expression
\begin{equation}
\frac{\bar n_m^0+1}{\bar n_m^0}= \frac{(\bar {\Delta} + \Omega_m)^2+\kappa^2/4} {(\bar {\Delta} - \Omega_m)^2+\kappa^2/4} \equiv \exp\left ( \frac{\hbar \omega_m}{k_B T_{\rm eff}}\right ), 
\end{equation}
$\bar n_m^T$ is the equilibrium phonon occupation determined by the mechanical bath temperature, and
\begin{eqnarray}
\label{gamma opt}
\Gamma_{\rm opt}&=&\frac{ \hbar g_0^2\langle \hat a^\dagger \hat a \rangle }{2m \Omega_m}  \\
&\times& \left [ \frac{\kappa}{(\bar {\Delta} + \Omega_m)^2+\kappa^2/4} - \frac{\kappa}{(\bar {\Delta} - \Omega_m)^2+\kappa^2/4}  \right ].\nonumber
\end{eqnarray}

For $\bar n_m^T \gg 0$ one recovers the classical result of Eq.~(\ref{cold-damping}). If the optical damping $\Gamma_{\rm opt}$ dominates,  $\Gamma_{\rm opt} \gg \gamma$, though, the mean phonon number is limited in the resolved sideband limit $\Omega_m \gg \kappa$  to
\begin{equation}
\bar n_m^0=\left (\frac{\kappa}{4 \Omega_m} \right )^2,
\label{fund limit}
\end{equation}
which shows that the ground state can be approached, but not reached, in that case. As expected from the classical considerations of the preceding section, this is the best possible case. In practice, the theoretical limit~(\ref{fund limit}) is difficult to reach due to technical noise issues including laser noise~\cite{RablGenes09,Phelps11}, clamping noise~\cite{Wilson08}, etc. but the discussion of these topics in beyond the scope of this brief review. Remarkably though, these experimental challenges have now being overcome in several experiments, see section 2.4. We also note that using pulsed optomechanical interactions may lead to improved cooling limits~\cite{Wang11,Machnes12}. We return briefly to this point in section 3.7.

Importantly, we remark that optomechanical sideband cooling is formally identical to the cooling of harmonically trapped ions, or more generally of any harmonically trapped dipole, see Ref.~\cite{Schliesser08} for a nice discussion of this point. In the case of trapped ions, the resolved sideband cooling limit was understood as early as 1975~\cite{Wineland,Neuhauser}, and the ground state cooling of trapped ions was first demonstrated over 20 years ago~\cite{Diedrich,Monroe}. As already mentioned, the key new element contributed by cavity optomechanics is the use of engineered resonance-enhancing structures.

\begin{figure}[t]
\includegraphics[width=0.45\textwidth]{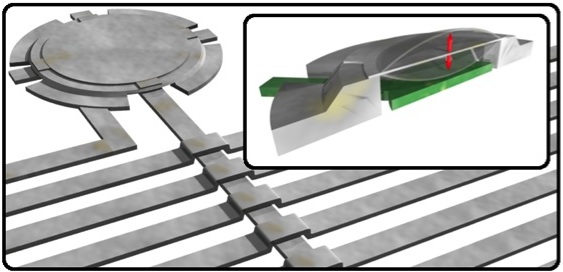}
\caption{(Color online) Artist conception of the microwave optomechanical circuit of Ref.~\cite{NIST}. Capacitor element of the LC circuit is formed by a 15 micrometer diameter membrane lithographically suspended 50 nanometers above a lower electrode. Insert: cut through the capacitor showing the membrane oscillations. After Ref.~\cite{Meystre11}.}
\end{figure}
\begin{figure}[t]
\includegraphics[width=0.5\textwidth]{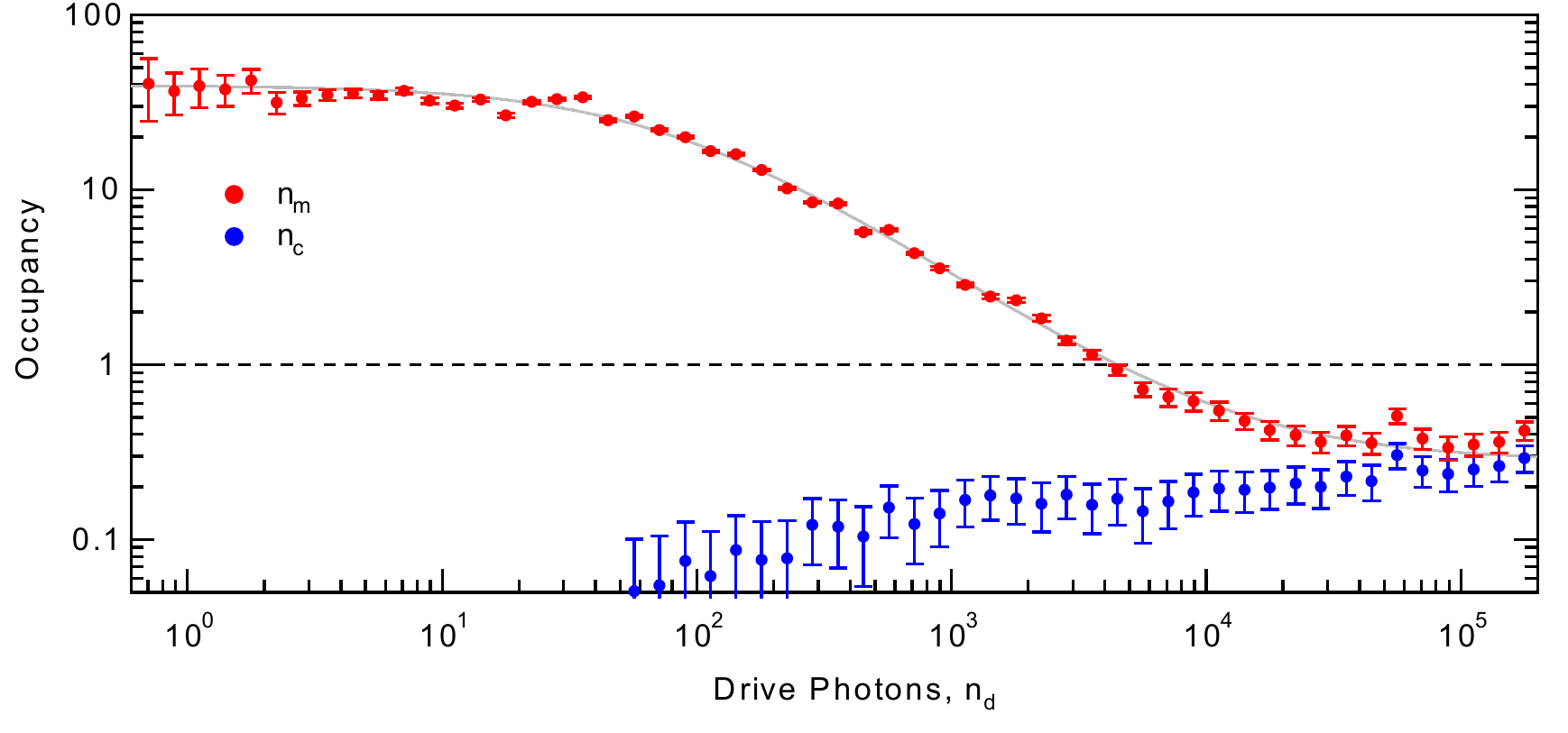}
\caption{(Color online)  Phonon occupancy (blue) and intracavity photon occupancy (red) as a function of the drive photon number. In this example sideband cooling reduces the thermal occupancy of the mechanical mode from  $n_m$=40 into the quantum regime, reaching a minimum of  $n_m$=0.34 $\pm$ 0.05. From Ref.~\cite{NIST}, with permission.}
\end{figure}

\subsection{Experimental status}

Following the pioneering work on gravitational wave antennas, advances in material science and nanofabrication -- in particular in microelectromechanical systems (MEMS), nanoelectromechanical systems (NEMS), and optical microcavities -- opened up the possibility to extend these ideas in many new directions, leading to the demonstration of significant cooling in a broad variety of systems from 2006 on, see Refs.~\cite{Arcizet06,Kleckner06,Schliesser06,Corbitt07}, with the first demonstration of cooling in the resolved sideband regime reported in Ref.~\cite{Schliesser08}. 

More recently these efforts have culminated in the cooling of the center-of-mass motion of at least three different micromechanical systems with a mean phonon number within a fraction of a phonon of their ground state of vibrational motion, $\langle n_m\rangle < 1$~\cite{UCSB, NIST, Caltech}. We postpone a discussion of Ref.~\cite{UCSB} until the next section to concentrate first on the two experiments \cite{NIST,Caltech} that utilized resolved sideband cooling to approach the mechanical ground state of center-of-mass motion. In one case~\cite{NIST} the mechanical resonator was a suspended circular aluminum membrane tightly coupled to a superconducting lithographic microwave cavity. That cavity was precooled to 20mK, corresponding to an initial occupation of 40 phonons and then further cooled by radiation pressure forces to an average phonon occupation of $\langle n_m\rangle \approx 0.3$. In contrast, Ref.~\cite{Caltech} utilized an optomechanical structure with co-located photonic and phononic band gaps in a suspended on-chip waveguide. The structure was precooled to 20K, corresponding to about 100 thermal quanta, and then cooled via radiation pressure to $\langle n_m\rangle \approx 0.85.$ Shortly thereafter, that same group also observed the motional sidebands generated on a second probe laser by a mechanical resonator cooled optically to near its vibrational ground state. They were able to detect the asymmetry in the sideband amplitudes between up-converted and down-converted photons, a smoking gun signature of the asymmetry between the quantum processes of emission and absorption of phonons~\cite{Painter12}.

\section{Beyond the ground state}

\subsection{Strong coupling regime}

Cooling mechanical resonators to their ground state of motion is an essential first step in eliminating the thermal fluctuations that normally mask quantum features. However, by itself that state is not particularly interesting, so the next challenge is to prepare, manipulate and characterize quantum states of the mechanical resonator required for a specific science or engineering goal. An important first experimental step in that direction was reported in Ref.~\cite{UCSB}. In contrast to Refs.~\cite{NIST} and \cite{Caltech} this experiment did not rely on radiation pressure cooling to achieve the motional ground state. Because of its high frequency of about 6 GHZ, a conventional dilution refrigerator that can reach temperatures of about 25 mK was sufficient to cool it to $\langle n_m\rangle <0.07$. A key point of the experiment is that it succeeded in coupling an acoustic resonator to a two-state system, or qubit, that could detect the presence of a single mechanical phonon. This is analogous to protocols that have been developed over the years in cavity quantum electrodynamics, see e.g. Ref.~\cite{HarocheBook}, with the important distinction that photons are now replaced by phonons. 

The coupling between a bosonic field mode and one or more two-state systems paves the way to a number of approaches to prepare and to observe genuine quantum features such as the energy quantization of the resonator, or to make controlled state manipulations at the few phonons level. Many of those protocols have already been developed in quantum optics and can be readily applied to phonon fields, at least in principle. In all cases dissipation and decoherence must be reduced to a minimum, as they rapidly lead to the destruction of the most salient quantum features of the state. In the experiment of Ref.~\cite{UCSB} decoherence was just weak enough to observe a few coherent oscillations of a single quantum exchanged between the qubit and the mechanical structure. As such it can be considered as the first demonstration of the capability of coherent control of phonon fields in a micromechanical resonator. 

Generally speaking, and in complete analogy with the situation in quantum optics and in cavity QED, the control of the quantum state of a mechanical oscillator requires that one operates in the so-called ``strong coupling regime,'' where the energy exchange between the mechanical object and the system to which it is coupled -- an optical field mode, a qubit, an electron, etc. -- is not negatively affected by dissipation and decoherence. Section~\ref{tql} showed that at the simplest level the optomechanical interaction takes the form (\ref{H opto}),
\begin{equation}
H=\hbar \omega_c \hat a^\dagger \hat a + \frac{\hat p^2}{2m} + \frac{1}{2} m \Omega_m^2 \hat x^2 -\hbar g_0 \hat a^\dagger \hat a (\hat b + \hat b^\dagger).
\label{H opto 2}
\end{equation}
At the single photon level, $\langle \hat a^\dagger \hat a \rangle =1$, this interaction is usually much too weak for its coherent nature to dominate over the incoherent dynamics for realizable levels of decay and decoherence. Since for $\langle \hat a^\dagger \hat a\rangle \gg 1$ the quantum nature of the optical field normally rapidly decreases in importance, it is therefore challenging to reach situations where the full quantum nature of the interaction between the photon and phonon fields is significant. There is  a way around this difficulty, though, the trade-off being that the intrinsic nonlinear nature of the optomechanical interaction  (\ref{H opto 2}) disappears in the process to be replaced by a linear effective interaction. As we shall see, this is not all bad, as that effective interaction offers itself a number of new opportunities.

Our starting point is the observation that strong intracavity optical fields can usually be decomposed as the sum of a classical, or mean-field part and a small quantum mechanical component. In terms of the mean field of the optical field mode $\alpha =\langle \hat a \rangle$
\begin{equation}
\hat a \rightarrow \alpha + \hat c
\end{equation}
 where $\hat c$ is again a photon annihilation operator. The optomechanical coupling term in the Hamiltonian (\ref{H opto 2}) becomes then
 \begin{equation}
 H_{\rm int}= -\hbar g_0 n (\hat b + \hat b^\dagger) -\hbar g \left ( \hat c + \hat c^\dagger \right )(\hat b + \hat b^\dagger)
\label{HBS}
 \end{equation}
 where we have introduced the optomechanical coupling strength
 \begin{equation}
 g = g_0 \sqrt{n},
\label{enhanced coupling}
 \end{equation}
$n = |\alpha|^2$, and we have taken $\alpha$ to be real for notational convenience. The first term in the Hamiltonian (\ref{HBS}) describes a simple Kerr effect, with a change in resonator length proportional to the classically intracavity intensity. This is the term that leads to the radiation pressure induced optical bistability observed e.g. in the experiments of Dorsel {\it et al}~\cite{Dorsel}. 

In a frame rotating at the driving field frequency, the cavity frequency and the mechanical frequency the second term in Eq.~(\ref{HBS}) can be reexpressed as
\begin{eqnarray}
V=&-&\hbar g \left [\hat b \hat c^\dagger e^{-i(\Delta + \Omega_m)t} + {\rm h.c.}\right ] \nonumber\\
&-&\hbar g \left [ \hat b^\dagger \hat c^\dagger e^{-i(\Delta - \Omega_m)t} + {\rm h.c.}\right ]
\label{strong coupling H}
\end{eqnarray}
This interaction describes the linear coupling between the quantized component of the optical field and the mechanical oscillator. The coupling $g$ is enhanced from the single-photon optomechanical coupling frequency $g_0$ by a factor $\sqrt{n}$, which can be very substantial. Note however that this enhancement comes at the cost of losing the nonlinear character of the original interaction $\hbar g_0 \hat a^\dagger \hat a (\hat b + \hat b^\dagger)$.  That nonlinear character is at the origin of a number of quantum effects that are expected to appear when the radiation pressure of a single (or of very few) photons displaces the mechanical oscillator by more than $x_{\rm zpf}$. These include two-photon blockade as well as quantitative changes in the output spectrum and cavity response of the optomechanical system, leading for example to the possible generation of non-Gaussian steady states of the oscillator \cite{Akram10,Rabl11,Nunnenkamp}. 

The linear coupling of Eq.~(\ref{strong coupling H}) provides exciting opportunities as well, and these are significantly less challenging to realize experimentally. On the red-detuned side of the Fabry-P{\'e}rot resonance, $\Delta = -\Omega_m$, we have after invoking the rotating wave approximation
\begin{equation}
 V \simeq -\hbar g \left (\hat b \hat c^\dagger  + {\rm h.c.} \right ),
\label{beam splitter H}
 \end{equation}
 the so-called beam-splitter Hamiltonian of quantum optics. In contrast, in the blue-detuned side of the resonance, $\Delta = +\Omega_m$, we have
\begin{equation}
 V\simeq -\hbar g \left (\hat b^\dagger \hat c^\dagger  + {\rm h.c.} \right ),
\label{squeezing H} 
\end{equation}
which describes the parametric amplification of the phonon mode and the optical field. 
 
This approach has enabled experiments to reach the regime of strong phonon-photon optomechanical coupling in several micromechanical devices~\cite{Groblacher09,Teufel11,Verhagen12}.  A familiar characteristic of strongly coupled systems is the occurrence of normal mode splitting. For the Hamiltonian~(\ref{strong coupling H}) the normal mode frequencies are
 \begin{equation}
 \omega_\pm=\frac{1}{2} \left [\Delta^2 + \Omega_m^2 \pm \sqrt{\left(\Delta^2-\Omega_m^2\right )^2 + 4g^2 \Omega_m \Delta} \right ]^{1/2}.
 \end{equation}
The first demonstration of normal mode coupling in an optomechanical situation was realized by Gr\"oblacher and coworkers~\cite{Groblacher09}. As pointed out by these authors the optomechanical modes can be interpreted in a dressed state approach as excitations of mechanical states that are dressed by the cavity radiation field. Alternatively, they can also be interpreted as optomechanical polariton modes.  Teufel and coworkers~\cite{Teufel11} carried a series of experiments in the strong coupling regime of quantum optomechanics. They  measured the dressed cavity states as a function of the pump-probe experiment where the coupling~(\ref{enhanced coupling}) was controlled by a pump field and the resonator transmission measured by a weak probe field. Increasing the strength of $g$ allowed them to monitor the change in cavity transmission as the strong coupling regime was reached, with an intermediate regime where the interference between the pump and probe field results in an effect analogous to electromagnetically induced transparenty~\cite{Boller91,Weis10}.
 \begin{figure}[t]
\includegraphics[width=0.45\textwidth]{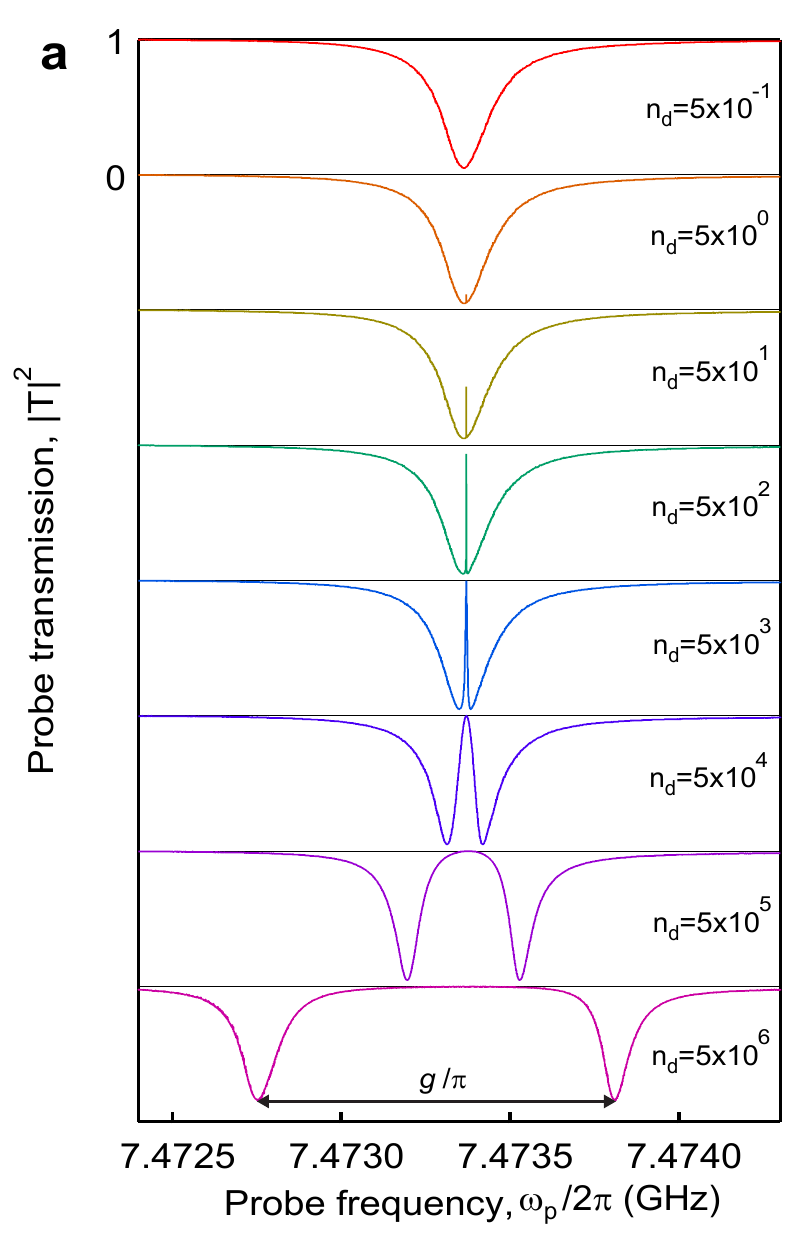}
\caption{(Color online) Normalized cavity transmission for increasing resonator drive intensity $n_d=|\alpha|^2$. For moderate drive intensities the interference between the drive and probe photons results in a narrow peak in the cavity spectrum, the onset of electromechanically induced transparency. For higher intensities the cavity resonance then splits into normal modes. From Ref.~\cite{Teufel11}, with permission.}
\end{figure}
 
 It should be emphasized that by itself, the observation of normal mode splitting, which is both a classical and quantum feature of coupled systems, does not prove the existence of coherent exchange of excitations between the mechanical and optical field modes. An important step toward the demonstration of quantum coherent coupling was recently achieved by Kippenberg and coworkers~\cite{Verhagen12} in a system where the optomechanical coupling is described by the beam splitter interaction~(\ref{beam splitter H}). This experiment considered a micro-mechanical oscillator cooled to a mean phonon number of the order of $\langle n_m\rangle \approx 1.7$, and in addition excited the system with a weak classical light pulses to achieve coherent coupling between the optical field and the micromechanical oscillator and the level of less than one quantum on average. These results, while still preliminary in many ways, open up a promising route towards the use of mechanical oscillators as quantum transducers, as well as in microwave-to-optical quantum links as we now discuss.

\subsection{State transfer}

The beam-splitter Hamiltonian (\ref{beam splitter H}) describes the coherent exchange of cavity photons and mechanical phonons. One of its remarkable properties is that it offers the potential to precisely transfer the quantum state of the mechanical oscillator to the electromagnetic field, and conversely. This is seen easily by considering the Heisenberg equations of motion for the annihilation operators $\hat b$ and $\hat c$ in the absence of decay,
\begin{eqnarray}
\hat b(t) &=& \hat b(0) \cos ( \hbar g t)   + i \hat c(0)\sin(\hbar g t) , \nonumber \\
\hat c(t) &=& \hat c(0) \cos ( \hbar g t)   + i  \hat b(0)  \sin( \hbar g t).
\end{eqnarray}
The optomechanical interaction $g$ can easily be made time dependent by pulsing the classical driving laser field intensity, $n \rightarrow n(t)$. For an interaction time $t_{\rm int}$ and a driving laser pulse intensity such that 
$$
\hbar g_0 \int_0^{t_{\rm int}}  dt \sqrt{n(t)} t = \pi/2
$$ 
we then have that $\hat b(t_{\rm int}) = \hat c(0)$ and $\hat c(t_{\rm int}) = i \hat b(0)$, indicative of a perfect state transfer between the optical and phonon modes -- assuming of course that dissipation and decoherence can be ignored during that time interval.

The interest in devices capable of high-fidelity state transfer between optical and acoustical fields is largely motivated by its potential for quantum information applications. This is because due to their potentially slow decoherence rate motional states of mechanical systems are well suited for information storage. However mechanics does not permit fast information transfer, while optical fields are ideal as information carriers, but are typically subject to fast decoherence that limits their interest for storage~\cite{Parkins}. The coherent quantum mapping of phonon fields to optical modes also promises to be useful in quantum sensing applications, by combining the remarkable sensitivity of nanoscale cantilevers to feeble forces and fields with reliable and high-efficiency optical detection schemes. And in addition to standard state transfer between motional and optical states,  phonon fields could also serve as convenient transducers between optical fields of different wavelengths, or between optical and microwave fields.  

The first theoretical proposal that analyzed a scheme to transfer quantum states from a propagating light field to the vibrational state of a movable mirror by exploiting radiation pressure effects is due to Jin Zhang and coworkers~\cite{ZhangBraunstein}. This work was then expanded in several directions, especially in the context of quantum optomechanics. For instance Tian and Wang~\cite{TianWang} proposed an optomechanical interface that converts quantum states between optical field of distinct wavelengths through a sequence of optomechanical $\pi/2$ pulses. In another recent proposal, Didier {\it et al.}~\cite{Didier} considered exploiting the beam splitter coupling of a mechanical oscillator and a microwave resonator to measure and synthesize quantum phonon states, and also to generate and detect entanglement between phonons and photons. They also proposed generating the entanglement of two mechanical oscillators and its detection by the cavity field after entanglement swapping. The first experimental demonstration of state transfer between a microwave field and a mechanical oscillator with amplitude at the single quantum level was recently achieved by Palomaki {\it et al.}~\cite{Palomaki}.
 
\subsection{Two-mode squeezing}

The Hamiltonian~(\ref{squeezing H}) is essentially the familiar two-mode squeezing Hamiltonian of quantum optics. This becomes readily apparent if one accounts for the (controllable) phase $\phi$ of the classical driving field, so that 
$$
g=g_0 \sqrt{n} \rightarrow i g_0 \sqrt{n}\exp(i\phi)
$$ 
and 
\begin{equation}
V=-i\hbar \left [g \hat b^\dagger \hat c^\dagger -  g^* \hat b \hat c\right ]
\end{equation}
with the associated evolution operator
\begin{equation}
S_{ab}(t)=\exp[(g^*\hat b \hat c - g\hat b^\dagger \hat c^\dagger)t ],
\end{equation}
the well-known unitary two-mode squeezing operator. Introducing the generalized two-mode quadrature operator
\begin{equation}
\hat X_{ab}= \frac{1}{2^{3/2}} (\hat c + \hat c^\dagger + \hat b + \hat b^\dagger) 
\end{equation}
one finds that the variance of a system initially in a two-mode vacuum state is given by~\cite{KnightLoudon}
 \begin{equation}
 \langle (\Delta X)^2\rangle = \frac{1}{4} \left [e^{-2|g|t}\cos^2(\phi/2)+ e^{2|g|t} \sin^2(\phi/2)\right ].
\end{equation}
That same result also holds if the two modes are initially in coherent states. For the choice $\phi = \pi/2$ one finds immediately that $\langle (\Delta X)^2\rangle$ can be well below the standard quantum limit of 1/4, a signature of two-mode squeezing. Two-mode squeezed states are known to be entangled, indicating that this form of interaction can result in quantum entanglement between the photon and phonon modes. As such this configuration represents a useful resource for demonstrating fundamental quantum mechanical effects as well as for exploiting cavity optomechanical devices in a quantum information context.

We note for completeness that in early work, Fabre and coworkers~\cite{Fabre94}, and independently Mancini and Tombesi~\cite{Mancini94} exploited the analogy between the situation of an optical resonator and a cavity filled by a Kerr medium to predict single mode squeezing of the reflected optical field in situations where the motion of the mirror is dominated by thermal fluctuations and can be treated classically. 
 
\subsection{Squeezing via back-action evading measurements}

As shown by Braginsky {\it et al.}~\cite{Braginsky80} and further analyzed by Clerk {\it et al.}~\cite{NJP-clerk} it is possible to implement back-action evading measurements of the membrane position when driving it with an input field resonant with the cavity frequency $\omega_c$, but modulated at the mirror frequency $\Omega_m$. The mean-field amplitude of the intracavity field is then 
\begin{equation}
\alpha(t) = \sqrt{n} \cos (\Omega_m t),
\end{equation}
where 
\begin{equation}
\alpha=\frac{\sqrt{\kappa}\alpha_{\rm in}}{i\Omega_m + \kappa/2}.
\end{equation}
Introducing the quadratures
\begin{eqnarray}
\hat X(t)&=& \frac{1}{\sqrt{2}} \left ( \hat ce^{i\Omega_m t} + \hat c^\dagger e^{-i\Omega_m t} \right ), \nonumber \\
\hat Y(t)&=& -\frac{i}{\sqrt{2}} \left ( \hat ce^{i\Omega_m t} - \hat c^\dagger e^{-i\Omega_m t} \right )
\end{eqnarray}
of the motional mode, with $[\hat X(t), \hat Y(t)]= i$ and 
\begin{equation}
\hat x(t) =\sqrt{2} x_{\rm zpt} \left (\hat X(t) \cos\Omega_m t + \hat Y(t)  \sin \Omega_m t \right ),
\end{equation}
and keeping  as before only linear terms in the quantum component of the field, the optomechanical interaction Hamiltonian reduces then to
\begin{equation}
V =-\sqrt{2}\hbar g \left[\hat X(1+\cos(2\Omega_m t))+\hat Y\sin(2\Omega_m t)\right]  (\hat b+\hat b^\dagger)
\label{intham1}
\end{equation}
where $g = g_0 \alpha = g_0 \sqrt{n}$ as before.

In a time-averaged sense the interaction Hamiltonian~(\ref{intham1}) reduces to
\begin{equation}
V  \rightarrow -\sqrt{2}\hbar g\hat X  (\hat b+\hat b^\dagger)
\label{time averaged H}
\end{equation}
and commutes with $\hat X$, thus giving rise to the possibility of performing a back action evading measurement of the $\hat X$ quadrature of mirror motion. This was verified experimentally in the classical regime by J.~B.~Hertzberg {\it et al.}~\cite{Hertzberg10}, but not yet in the quantum regime so far.

Since the interaction~(\ref{time averaged H}) is linear in $\hat X$ it is perhaps less evident that it can also lead to quadrature squeezing. This can be achieved by first performing a precise measurement of $\hat X$, following which its quadrature can clearly be below the standard quantum limit. Following that measurement the system would normally rapidly relax back to a classical state, but by applying an appropriate feedback, the measurement induced squeezing can be turned into real squeezing. This is discussed in detail in Ref.~\cite{NJP-clerk}.

\subsection{Parametric instability}

We have seen that for a driving laser red-detuned from the cavity frequency $\omega_c$ the upper sideband is resonantly enhanced by the cavity, which leads to preferred extraction of mechanical energy, i.e. cavity cooling. For blue-detuned light, in contrast, it is the lower sideband that is resonantly enhanced by the cavity, resulting in the preferred deposition of mechanical energy, i.e. the optical amplification of mechanical motion. Invoking the rotating-wave approximation for $\Delta \approx +\Omega_m$ one finds that this process is described at the simplest level by the 2-mode squeezing interaction~(\ref{squeezing H}) instead of the beam-splitter Hamiltonian~(\ref{beam splitter H}) of section 3.2. 

In that regime the optomechanical system can display dynamical instabilities. For appropriate parameters they result in stable mechanical oscillations somewhat reminiscent of laser action, but for a phononic field~\cite{Vahala09,Grudinin10}, or even in unstable dynamics and chaos~\cite{Kippenberg07}. That this can be the case is already apparent at the classical level from the fact that $\Gamma_{\rm opt}$ can become negative for blue detuning, see Eq.~(\ref{gamma opt}). If the laser intensity is strong enough that the total damping rate $\gamma + \Gamma_{\rm opt}$ is itself negative, then any amplitude oscillation will grow exponentially until it saturates due to the onset of nonlinear effects.

Following Ludwig~{\it et al.}~\cite{Ludwig} we assume that the motion of the cantilever is approximately sinusoidal,
\begin{equation}
x(t) \approx \bar x + A \cos (\Omega_m t),
\label{x ansatz}
\end{equation}
with the average position $\bar x$ given by the radiation pressure force,
\begin{equation}
\bar x = \frac{1}{m\Omega_m^2} \langle F_{\rm rad} \rangle = \frac{\hbar G}{m\Omega_m^2} \langle |\alpha(t)|^2\rangle
\end{equation}
where $\langle \alpha(t)|^2\rangle$ is the intracavity light intensity and $A$ is the amplitude of oscillations of the mirror. Marquardt and coworkers~\cite{Marquardt06} showed that with Eq.~(\ref{x ansatz}), Eq.~(\ref{field eq}) yields for the intracavity field
\begin{equation}
\alpha(t) = e^{i\phi(t)} \sum_n\alpha_n e^{i n \Omega_m t},
\end{equation}
with
\begin{equation}
\alpha_n =\left (\frac{\alpha_{\rm max}}{2}\right )  \frac{J_n(-GA/\Omega_m)}{in\Omega_m/\kappa + i(G \bar x - \Delta)/\kappa + 1/2}.
\end{equation}
Here $\phi(t)=(GA/\Omega_m)\sin(\Omega_m t)$ and $J_n$ are Bessel functions of the first kind. The stability of the system can be determined simply comparing the mechanical power $P_{\rm rad}$ due to radiation pressure to the dissipated power $P_{\rm fric}$ due to friction. When their ratio increases above unity the system starts to undergo self-induced oscillations~\cite{Ludwig}. 

Figure~6 is an example of a stability diagram determined from such an analysis. It shows the ratio $P_{\rm rad}/P_{\rm fric}$ as a function of the detuning $\Delta$ and the square of the (dimensionless) mechanical energy $A^2$. Regions with $P_{\rm rad}/P_{\rm fric} > 1$ are unstable, and the solid line defines an attractor where there is an exact power balance between amplification and damping. In general the parameter space $(\Delta, A)$ is characterized by the presence of a number of such attractors. For relatively weak amplitudes $A$, nonlinear effects tend to stabilize the oscillations of the cantilever, leading to "laser-like" oscillations~\cite{Vahala09,Grudinin10}, but for larger oscillations amplitudes the system can become chaotic~\cite{Kippenberg07}.  Quantum mechanically fluctuations are strongly amplified just below threshold, so that the attractor is no longer sharp~\cite{Jiang11}. 

\begin{figure}[t]
\includegraphics[width=0.55\textwidth]{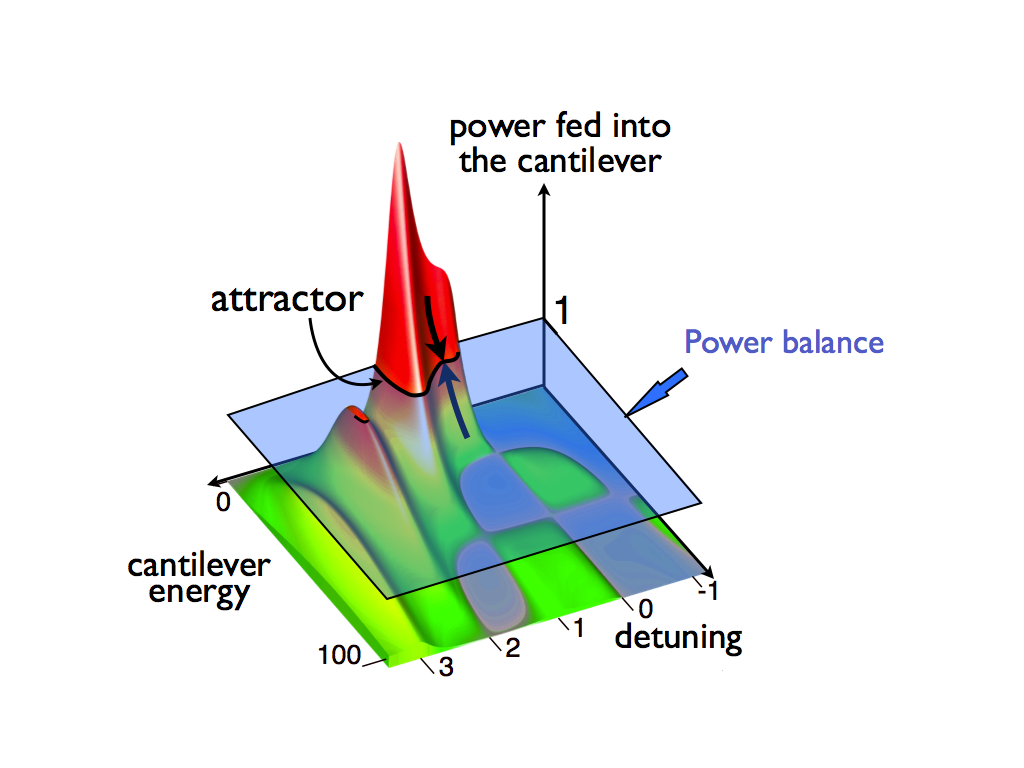}
\caption{(Color online) Attractor diagram obtained from the requirement that the optical power fed into mechanical oscillations is balanced by the power lost to friction. Adapted from From Ref.~\cite{Ludwig}, with permission.} 
\end{figure}

\subsection{Quadratic coupling}

So far we have considered geometries where the optomechanical coupling is linear in the oscillator displacement. Other forms of coupling can however be considered, most interestingly perhaps a coupling quadratic in the displacement. This can be realized in so-called ``membrane-in-the-middle'' geometries, as first demonstrated by J. Harris and his collaborators at Yale~\cite{Thompson08,Jayich08}, see also Refs.~\cite{Meystre85,Mishkat08}.  As implied by its name, this geometry involves an oscillating mechanical membrane placed inside a Fabry-P{\'e}rot with fixed end-mirrors. 

An attractive feature of membrane-in-the-middle configurations is the ability to realize relatively easily either linear or quadratic optomechanical couplings, depending on the precise equilibrium position of the membrane. In case the membrane is located at an extremum of $\omega_c'(x)$, so that  $G=-\partial \omega_c'/\partial x =0$, see Eq.~(\ref{G}), we have to lowest order
\begin{equation}
\omega_c'(x) \approx \omega_c + \frac{1}{2} \frac{\partial^2 \omega_c}{\partial x^2}
\end{equation}
so that the optomechanical Hamiltonian becomes
\begin{equation}
H=\hbar \omega_c \hat a^\dagger \hat a + \hbar \Omega_M \hat b^\dagger \hat b + \frac{1}{2} \frac{\partial^2 \omega_c}{\partial x^2} x_{\rm zpt}^2 (\hat b + \hat b^\dagger)^2 \hat a^\dagger \hat a.
\end{equation}
In the rotating wave approximation this reduces to
\begin{eqnarray}
H&=&\hbar \omega_c \hat a^\dagger \hat a + \hbar \Omega_M \hat b^\dagger \hat b +  \hbar x_{\rm zpf}^2\frac{\partial^2 \omega_c}{\partial x^2 } \left ( \hat b^\dagger \hat b+1/2\right ) \hat a^\dagger \hat a \nonumber \\
&=&\hbar \omega_c \hat a^\dagger \hat a + \hbar \Omega_M \hat b^\dagger \hat b +\hbar g_0^{(2)}  \left (\hat b^\dagger \hat b+1/2 \right ) \hat a^\dagger \hat a 
\end{eqnarray}
where
\begin{equation}
g_0^{(2)}\equiv x_{\rm zpf}^2 \frac{\partial^2 \omega_c}{\partial x^2} .
\end{equation}
Quadratic coupling opens up the way to a number of interesting possibilities, including the direct measurement of energy eigenstates of the mechanical element, rather than the position detection characteristic of linear coupling. J.~Harris and coworkers estimate that it may be possible in the future to use this scheme to observe quantum jumps of a mechanical system~\cite{Jayich08}. In another theoretical study, Nunnenkamp and coworkers~\cite{Nunnenkamp10} considered  optomechanical cooling and squeezing via quadratic optomechanical coupling. They showed that for high temperatures and weak coupling, the steady-state phonon number distribution is nonthermal, and demonstrated how to achieve mechanical squeezing by driving the cavity with two optical fields.

Another possibility offered by that geometry is to observe the quantum tunneling of an optomechanical system operating deep in the quantum regime through a classically forbidden potential barrier. One proposed approach~\cite{Buchmann12} relies on adiabatically raising a potential barrier, whose parameters can be widely tuned, at the location of a mechanical element. For the right choice of parameters the optomechanical potential is a double-well potential, and it is estimated that quantum tunneling between its wells can occur at rates several orders of magnitude larger than the decoherence rate of the mechanical membrane. Besides tunneling, that scheme may also allow for the study of the quantum Zeno effect in a mechanical context and provide a comparatively simple scheme for the preparation and characterization of non-classical mechanical states of interest for quantum metrology and sensing.  

\subsection{Pulsed optomechanics}

So far we have largely limited our discussion to situations where the optomechanical coupling is either constant or slowly varying in time. One notable exception was the quantum state transfer protocol outlined in Section III.B, which requires that the interaction $g(t)$ be turned off at the precise time when the state transfer has been completed. However there are a number of situations where pulsed interactions are desirable, as already realized by Braginsky~\cite{BraginskyBook,Braginsky78} in his proposal for a back-action evading position measurement scheme. In a recent paper, Vanner and coworkers~\cite{Vanner11} proposed to use a pulsed interaction of duration short compared to the period of the mechanical oscillator to generate and fully reconstruct quantum states of mechanical motion: As a result of the interaction the phase of the pulsed driving optical field becomes correlated with the position of the mechanical oscillator, while its intensity imparts it a momentum boost. A time domain homodyne detection scheme can then be used to measure the phase of the field emerging from the cavity, thereby providing a measurement of the mechanical position. This scheme can also be used to achieve squeezing and state purification of the mechanical resonator. It has also recently been proposed that pulsed optomechanics could be used, at least in principle, to surpass the limits of conventional sideband cooling by using an optimized sequence of driving optical pulses~\cite{Wang11,Machnes12}.

In an intriguing potential application of pulsed optomechanics, Pikovski and colleagues~\cite{Pikovski12} considered a scheme to measure the canonical commutator of a massive mechanical oscillator and by doing so to detect possible commutator deformations due to quantum gravity: there are speculations that the existence of a minimum length scale where space-time is assumed to be quantized, possibly of the order of the Planck length $L_P= 1.6 \times 10^{-35}$m, could result in such deformations. In this proposal a sequence of optomechanical interactions would be used to map the commutator of the mechanical resonator onto an optical pulse. Remarkably the analysis of Ref.~\cite{Pikovski12} suggests that as a result Planck-scale physics might be observable in a relatively mundane quantum optics experiment.

\section{Cold atoms}

In a development complementary to the research on nanoscale mechanical systems, recent quantum optomechanics experiments have also manipulated and controlled at the quantum level  the center-of-mass degrees of freedom of ultracold atomic ensembles~\cite{Murch08,Brennecke08,Kanamoto09,Schleier11}. In the following we restrict our discussion to the case of a neutral atomic sample cooled well below its recoil temperature and trapped inside  a single-mode Fabry-P{\'e}rot resonator. This could be for example a nearly homogeneous and collisionless Bose-Einstein condensate (BEC) at $T \approx 0$ or a sample cooled near the vibrational ground state of one or a few wells of the optical lattice formed by the optical field. Side mode excitations of the condensate in the first case, and the vibrational motion of thermal atoms in the second case, provide formal analogs of one or several moving mirrors. 

To see how this works we consider first a generic model consisting of a BEC at $T=0$ trapped inside a Fabry-P\'erot cavity of length $L$ and mode frequency $\omega_c$. The atoms of mass $M$ are driven by a pump laser of frequency $\omega_L$ and wave number $k$. When $\omega_L$ is far detuned from the atomic transition frequency $\omega_a$ the excited electronic state of the atoms can be adiabatically eliminated and the atoms interact dispersively with the cavity field. In the dipole and rotating-wave approximations, the Hamiltonian describing the interaction between the atoms and the optical field is 
\begin{equation}
H=H_{\rm atom} + H_{\rm field}, 
\end{equation}
where
\begin{equation}
H_{\rm field}=\hbar \omega_c \hat a^\dagger \hat a
\end{equation} 
and
\begin{equation}
H_{\rm bec}=\int dx \hat{\Psi}^{\dagger}(x) \left[\frac{\hat{p}_x^2}{2M}+\hbar U_0\cos^2 (kx)\hat a^{\dagger}\hat a\right] \hat{\Psi}(x).
 \label{Hatom}
\end{equation}
Here $\hat \Psi(x)$ is the bosonic Schr\"odinger field operator for the atoms, $\hat a$ is the photon annihilation operator as before, and the atoms interact with the light field via the familiar off-resonant coupling 
\begin{equation}
U_0=g_R^2/(\omega_L-\omega_a),
\label{Rabi}
\end{equation}
where $g_R$ is the single-photon Rabi frequency. As always $H$ should be complemented by contributions describing the external driving of the cavity field, dissipation and collisions.

When the light field can be approximated as a plane wave the atomic field operator can likewise be expanded in terms of plane waves as
\begin{equation}
\hat{\Psi}(x) =( 1/\sqrt{L})\sum_{q} \hat b_k e^{iqx},
\end{equation} 
where $\hat{b}_q$ and $\hat{b}_q^{\dagger}$ are annihilation and creation operators for atomic bosons with the momentum $k$, satisfying the bosonic commutation relations $[\hat{b}_q,\hat{b}_{q'}^{\dagger}]=\delta_{q,q'}$ and $[\hat{b}_q,\hat{b}_{q'}]=0$. 

Consider for simplicity the case of scalar bosonic atoms:  In the absence of light field and at $T=0$ the ground state of the sample would be a condensate with zero momentum, 
\begin{equation}
|\Psi_0\rangle= (\hat{b}_{0}^{\dagger})^N|0\rangle,
\end{equation}
but as a result of virtual transitions the atoms can acquire a recoil momentum $\pm 2\ell \hbar k$, where $\ell = 0, 1, 2, \ldots$ In the limit of low photon numbers it is sufficient to consider the lowest diffraction order, $\ell = 1$ and the atomic field operator can be conveniently expressed in terms of a zero-momentum component and a ``sine mode,''
\begin{eqnarray}\label{B_fo}
\hat{\Psi} \sim \hat{b}_0 \phi_0 (x) + \hat{b}_2 \phi_2 (x)
\label{mode decomp}
\end{eqnarray}
where $\phi_0(x)$ is the condensate wave function and $\phi_2 (x)=\sqrt{2}\cos(2kx)$. For very weak optical fields the occupation of the sine mode remains much smaller the the zero-momentum mode,  so that $\hat{b}_0 \simeq \sqrt{N}$ and $\langle \hat{b}_2^{\dagger}\hat{b}_2\rangle \ll N. $ Substituting then Eq.~(\ref{B_fo}) into the Hamiltonian~(\ref{Hatom}) and in a frame rotating at the pump laser frequency the Hamiltonian $\hat{H}$ becomes
\begin{equation}
\label{HeffB}
\hat{H}_{\rm om,bec}=4\hbar \omega_{\rm rec}\hat{b}_2^{\dagger}\hat{b}_2 + \hbar\hat{a}^{\dagger}\hat{a} \left[\Delta+g_2(\hat{b}_2+\hat{b}_2^{\dagger})\right],
\end{equation}
where 
\begin{equation}
g_2=(U_0/2)\sqrt{N/2}
\end{equation}
is the effective atom-field coupling constant, $\omega_{\rm rec}=\hbar K^2/2M$ the recoil frequency, and $\Delta=\omega_c-\omega_L+U_0 N/2$ is an effective Stark-shifted detuning .

The reduced Hamiltonian~(\ref{HeffB}) describes the coupling of two oscillators, the cavity mode $\hat a$ and the momentum side mode $\hat b_2$ via the optomechanical coupling $g_2\hat a^\dagger \hat a(\hat b_2+\hat b_2^\dagger )$. This shows that the condensate momentum side mode behaves formally like a moving mirror driven by the radiation pressure of the intracavity field, see Eq.~(\ref{H quantum opto}) for comparison.

A similar analogy can be established when considering a sample of ultracold atoms tightly confined to an harmonic trap of frequency $\omega_z$  centered at some location $z_0$ along the resonator axis. The position of atom $i$ is then $z_i = z_0 +\delta z_i$, and the vacuum Rabi frequency with which it interacts with the field is 
\begin{equation}
g_R(z_i) = g_R \sin(\phi_0 + 2k \delta z_i),
\end{equation}
where $\phi_0=k z_0$
so that Eq.~(\ref{Rabi}) becomes
\begin{equation}
U_0=\frac {g(z_i)^2}{\omega_L-\omega_a}.
\end{equation}
Summing over all atoms in the sample and expanding then the far off-resonant atom-field interaction to lowest order in $K \delta z_i$ one finds for $ \omega_L = \omega_c$~\cite{Purdy10}
\begin{eqnarray}
H &\approx& \hbar (\omega_c + NU_0 \sin^2 \phi_0) \hat a^\dagger \hat a + \hbar \omega_z \sum_i \hat b_i^\dagger \hat b_i \nonumber \\
&+& \hbar U_0 \sin (2\phi_0) \hat a^\dagger \hat a \left [\sum_i  k\delta z_i \right ]
\label{hatoms}
\end{eqnarray}
where $N$ is the number of atoms and the operator $\hat b_i$ describes the annihilation of a phonon from the center-of-mass motion of atom $i$. 

The second line of the Hamiltonian~(\ref{hatoms}) describes the optomechanical coupling of the intracavity optical field to the collective atomic variable 
\begin{equation}
 k\sum_i \delta z_i = kN Z_{\rm cm}
 \end{equation}
 which is nothing but a measure of the normal mode of the sample, its center of mass $Z_{\rm cm}= N^{-1}\sum_i \delta z_i$. For small displacements that mode can be described as a harmonic oscillator of frequency $\omega_z$ and mass $NM$. In this picture, the atom-field system is therefore modeled by the optomechanical Hamiltonian
\begin{equation}
H_{\rm om, at} =\hbar \omega_c' \hat a^\dagger \hat a + \hbar \omega_z \hat b^\dagger \hat b + \hbar g_N(\hat b + \hat b^\dagger)\hat a^\dagger \hat a,
\end{equation}
where $\hat b$ and $\hat b^\dagger$ are bosonic annihilation and creation operators for the center-of-mass mode of motion of the atomic ensemble, $z_{\rm zpf} = \sqrt{\hbar/2Nm \omega_z}$ and 
\begin{equation}
g_N=NU_0   (Kz_{\rm zpf}) \sin^2\phi_0
\end{equation}
with scales as $\sqrt{N}$. Quantum optomechanics experiments with non-degenerate ultracold atoms samples have so far been carried out principally in the group of D.~Stamper-Kurn at UC Berkeley, while T.~Esslinger and coworkers at ETH Z\"urich have concentrated on the use of Bose condensates~\cite{Stamper12}.  In a trailblazing experiment~\cite{Purdy10} Purdy {\it et al} positioned a sample of cold atoms with sub-wavelength accuracy in a Fabry-P{\'e}rot cavity to demonstrate the tuning from linear to quadratic optomechanical coupling from the linear to the quadratic coupling regime. The Berkeley group also observed the measurement back-action resulting from the quantum fluctuations of the optical field by measuring the cavity-light-induced heating of the atomic ensemble~\cite{Murch08}, the first observation of quantum back-action on a `macroscopic' mechanical resonator at the standard quantum limit. More recent work~\cite{Brahms12} detected the asymmetric coherent scattering of light by a collective mode of motion of a trapped ultracold gas with 0.5 phonons of average excitation, a result that complements the work of Safavi-Naeini {\it et al.}~\cite{Painter12} on the asymmetric absorption of light by a nanomechanical solid-state resonator, see section 2.4.

\begin{figure}[t]
\includegraphics[width=0.5\textwidth]{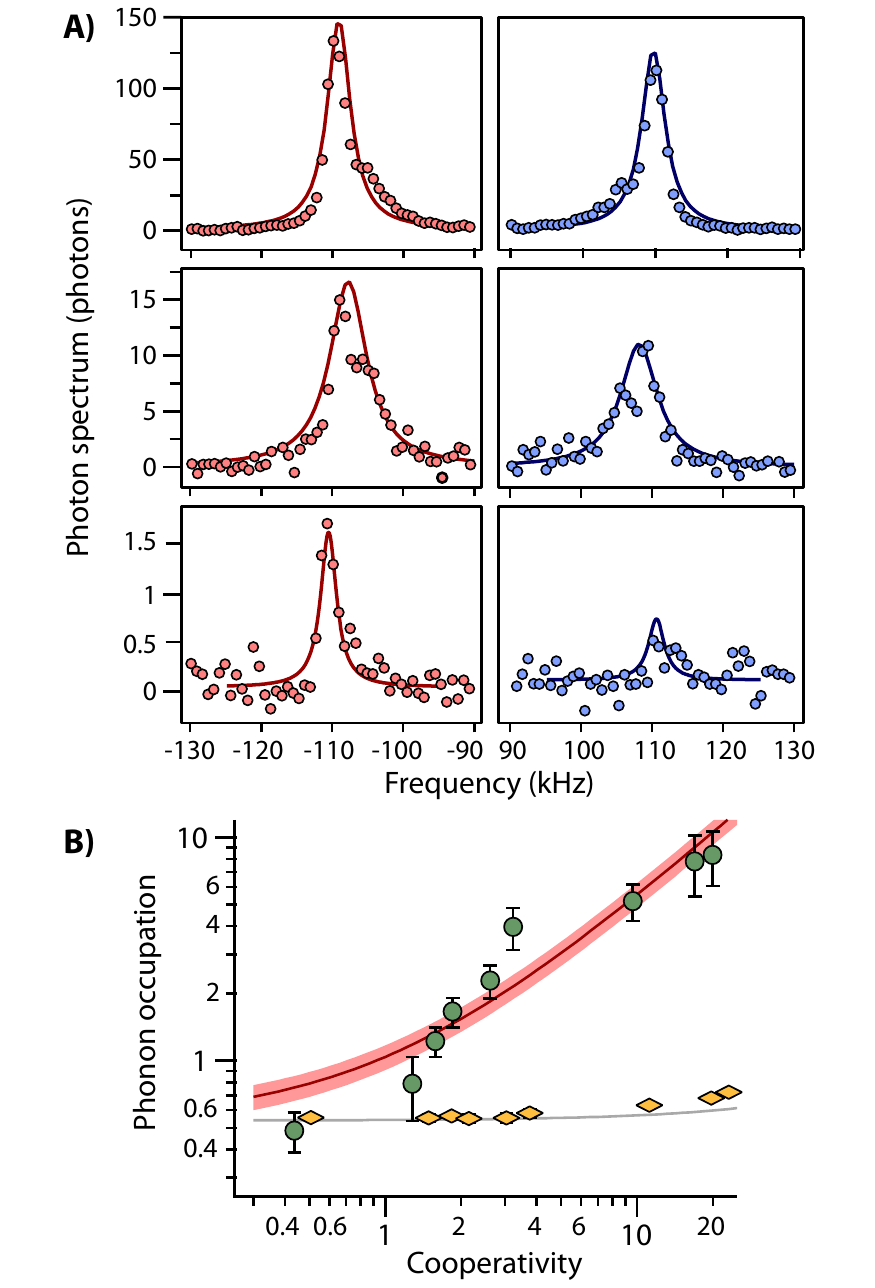}
\caption{(Color online)  (a) Asymmetric optical scattering from quantum collective motion, with the measured Stokes sidebands [left panels, (red) circles] and anti- Stokes sidebands [right panels, (blue) circles] at various mean phonon numbers, characterized by the so-called cooperatively coefficient $C$. From top to bottom $C$= 9.6; 1.9; 0.4.~(b)  Measured phonon occupation vs cooperativity. From Ref.~\cite{Brahms12}, with permission.}
\end{figure}

Turning now to quantum degenerate gases, Brenneke {\it et al.}~\cite{Brennecke08} studied the dynamics of a Bose condensate of $^{87}$Rb atoms trapped inside a high-finesse Fabry-P{\'e}rot and driven by a feeble optical field. This experiment demonstrated the optomechanical coupling of a collective density excitation of the condensate, showing that it behaves precisely as a mechanical oscillator coupled to the cavity field, in quantitative agreement with a cavity optomechanical model of Eq.~(\ref{HeffB}). These authors also succeeded in approaching the strong coupling regime of cavity optomechanics, where a single excitation of the mechanical oscillator substantially influences the cavity field.  In subsequent work, the Bose condensate was irradiated from the side of the optical resonator, resulting in the demonstration of a second-order quantum phase transition where the condensed atoms enter a self-organized super-solid phase, a process mathematically described by the Dicke model of an ensemble of two-state systems coupled to a single-mode electromagnetic field. In contrast to the situation in the usual Dicke model, where the two states of interest are two atomic electronic levels coupled by a dipole optical transition, in the present case the relevant states are two different momentum states coupled to the cavity field mode~\cite{Baumann10,Baumann11}.

\section{Outlook -- Functionalization and hybrid systems}

The rapid progress witnessed by quantum optomechanics makes it increasingly realistic to consider the use of mechanical systems operating in the quantum regime to make precise and accurate measurements of feeble forces and fields \cite{Kippenberg08}. In many cases, these measurements amount to the detection of exceedingly small displacements, and in that context the remarkable potential for functionalization of opto-- and electromechanical devices is particularly attractive. Their motional degree(s) of freedom can be coupled to a broad range of other physical systems, including photons via radiation pressure from a reflecting surface, spin(s) via coupling to a magnetic material, electric charges via the interaction with a conducting surface, etc. In that way, the mechanical element can serve as a universal transducer or intermediary that enables the coupling between otherwise incompatible systems. This potential for functionalization also suggests that quantum optomechanical systems have the potential to play an important role in classical and quantum information processing, where transduction between different information carrying physical systems is crucial.

Much potential for the functionalization of optomechanical devices is offered by interfacing them with a single quantum object. This could be an atom or a molecule, but also an artificial atom such as a nitrogen vacancy center (NV center) in diamond~\cite{Rabl09}, a superconducting qubit~\cite{Armour02,LaHaye09,UCSB} or a Bose-Einstein condensate~\cite{Treutlein07}.  Several theoretical proposals~\cite{Treutlein07,Genes08,Singh08,Geraci09,Hammerer09,Genes09,Rabl09,Hammerer10} and more recently experimental realizations~\cite{Hunger10,Treutlein11} involving atomic systems have been reported. For example, a recent experiment~\cite{Camerer11} realized a hybrid optomechanical system by coupling ultracold atoms trapped in an optical lattice to a micromechanical membrane, their coupling being mediated by the light field.  Both the effect of the membrane motion on the atoms and the back-action of the atomic motion on the membrane were observed. Singh and coworkers~\cite{Singh12} considered a variation on that scheme where a Bose condensate is trapped inside a Fabry-P{\'e}rot with a moveable end mirror driven by a feeble optical field. They showed that under conditions where the optical field can be adiabatically eliminated one can achieve high fidelity quantum state transfer between a momentum side mode of the condensate, see Eq.~(\ref{mode decomp}), and the oscillating end-mirror. 

Artificial atoms such as NV centers are of much interest for hybrid optomechanical systems~\cite{Rabl09} due to the attractive combination of their optical and electronic spin properties. Their ground state is a spin triplet~\cite{Doherty12} that can be optically initialized, manipulated and read-out by a combination of optical and microwave fields, and they are characterized by remarkably long room-temperature coherence times for solid-state systems. As such, they offer much promise for applications e.g. in quantum information processing and ultrasensitive magnetometry, where the spin is used as an atomic-sized magnetic sensor~\cite{Taylor08,Maze08,Bala08}. In this context, a spin-oscillator system of particular interest consists of a magnetized cantilever coupled to the electronic spin of the NV center. A recent experiment by Arcizet and colleagues demonstrated the coupling of a nanomechanical oscillator to such a defect in a diamond nanocrystal attached to its extremity~\cite{Arcizet11}. 

In two further recent demonstrations of the potential of hybrid optomechanical systems, a mechanical oscillator was used to achieve the coherent quantum control of the spin of a single NV center~\cite{Hong12}, and the coherent evolution of the spin of an NV center was coupled to the motion of a magnetized mechanical resonator to sense its motion with a precision below 6 picometers~\cite{Kolkowitz12}. The authors of that experiment comment that it may soon become possible to detect the mechanical zero-point fluctuations of the oscillator.

More speculatively perhaps, micromechanical oscillators in the quantum regime offer a route toward new tests of quantum theory at unprecedented sizes and mass scales. For instance, spatial quantum superpositions of massive objects could be used to probe various theories of decoherence and shed new light on the transition from quantum to classical behavior: In contrast to the generally accepted view that it is technical issues such as environmental decoherence that rapidly destroy such superpositions in massive objects and establish the transition from the quantum to the classical world, some authors~\cite{Ghirardi86,Frenkel90,Penrose96,Diosi07,Ellis92} have proposed collapse models that are associated with more fundamental mechanisms and the appearance of new physical principles. Bouwmeester~\cite{Bouwmeester} has pioneered the idea that quantum optomechanics experiments may shed light on this issue and on possible unconventional decoherence processes, and in recent work Romero-Isart has analyzed the requirements to test some of these models and discussed the feasibility of a quantum optomechanical implementation using levitating dielectric nanospheres~\cite{Isart10,Isart12}.

\begin{acknowledgements}
This work is supported by the US National Science Foundation, by the DARPA ORCHID and QuASAR programs through grants from AFOSR and ARO,  and by the US Army Research Office. We acknowledge enlightening discussions with numerous colleagues, in particular M. Aspelmeyer, L. Buchmann, A. Clerk, H. Jing, M. Lukin, R. Kanamoto, K. Schwab, H. Seok, S. Singh, S. Steinke, M. Vengalattore, E. M. Wright and K. Zhang.
\end{acknowledgements}

\end{document}